\documentclass[]{aiaa-tc}
\usepackage{booktabs,amsmath,amsthm,amssymb,amsbsy,latexsym,graphicx,color}
\usepackage{rotating}
\usepackage{float}
\usepackage{subfigure}
\usepackage{upgreek}
\usepackage{bm}
\usepackage{mathdots}
\usepackage{relsize}
\usepackage{caption}
\usepackage{newfloat}
\usepackage{fixltx2e}
\usepackage{nomencl}
\usepackage{multicol}
\usepackage{indentfirst}
\usepackage{etoolbox}
\usepackage{tocvsec2}
\usepackage{gensymb}
\usepackage[titletoc]{appendix}
\usepackage{appendix}
\usepackage[bbgreekl]{mathbbol}
\usepackage{graphicx}

\title{A Three-Dimensional Constitutive Modeling for Shape Memory Alloys Considering Two-Way Shape Memory Effect and Transformation-Induced Plasticity}

 \author
  {
  	Lei Xu
  	\thanks{Graduate Research Assistant, Aerospace Engineering, College Station, TX 77843, USA, Student member.}\thanksibid{1}\\
  {\normalsize\itshape
   Texas A\&M University, College Station, TX 77843, USA}\\
  \and
  Alexandros Solomou
  \thanks{Post-doctoral Research Associate, Aerospace Engineering, College Station, TX 77843, USA}\thanksibid{2}\\
  {\normalsize\itshape
	 Texas A\&M University, College Station, TX 77843, USA}\\
   \and
   Dimitris Lagoudas
   \thanks{ Distinguished Professor,Aerospace Engineering,College Station, TX 77843, USA, Faculty member.}\thanksibid{1}\\
  {\normalsize\itshape
  Texas A\&M University, College Station, TX 77843, USA}\\
  }

\begin{document}

\maketitle

\begin{abstract}
Shape memory alloys (SMAs), since the discovery of their shape memory effect, have been intensively investigated as actuators for the past several decades. Due to their high actuation energy density compared to other active materials, their current and potential applications in the biomedical, aerospace, automobile and energy fields are expanding rapidly. Prior to be used as actuators, SMAs are usually subjected to a training process (i.e. thermal cycling under isobaric conditions) to stabilize their behavior. During the training process, permanent changes are introduced in the microstructure of the material which results in the generation of internal stresses and a large amount of irrecoverable Transformation Induced Plastic strain (TRIP). The generated internal stresses along with a potential thermal loading provide the driving force to induce the oriented phase transformation so that the SMA-based actuators are able to exhibit the Two-Way Shape Memory Effect (TWSME) without applying external bias load. To predict this intrinsic phenomenon, a three-dimensional phenomenological constitutive model for untrained SMAs is presented. The proposed model utilizes the martensitic volume fraction, transformation strain, TRIP strain, and internal stress as internal state variables so that it is able to account for the evolution of TRIP strain and the TWSME for untrained SMAs under cyclic thermomechanical loading conditions. In the end, boundary value problems considering an untrained SMA material under isothermal/isobaric cyclic loading are solved and the predicted cyclic response is compared against available experimental data to demonstrate the proposed capabilities. The proposed model is anticipated to be further validated against additional experimental data for NiTiHf SMAs under general 3-D loading conditions. The validated model will be utilized as an efficient tool for the design and analysis of SMA-based actuators, such as SMA beams and torque tubes, which are intended for the realization of the future supersonic transport aircraft with morphing capabilities to reduce the sonic boom noise.

\end{abstract}

\newpage
\section*{Nomenclature}

\begin{multicols}{2}
	\begin{tabbing}
		\bfseries left \quad \=\bfseries center \quad \=\bfseries right \quad \=\bfseries paragraph \kill

		$\mathcal{S}$ \>\quad Forth order compliance tensor\\  
		$\mathcal{S}^A$ \>\quad Forth order compliance tensor of austenite\\  
		$\mathcal{S}^M$ \>\quad Forth order compliance tensor of martensite\\ 
		$\Delta \mathcal{S}$ \>\quad Phase difference of compliance tensor \\
		\>\quad between austenite and martensite \\   
		$\bm\Lambda$ \>\quad Transformation direction tensor\\
		$\bm\Lambda_{fwd}$ \>\quad Forward transformation direction tensor\\
		$\bm\Lambda_{rev}$ \>\quad Reverse transformation direction tensor\\
		$\bm\Lambda^p$ \>\quad TRIP strain direction tensor\\	
		$\bm\Lambda^p_{fwd}$ \>\quad Forward TRIP strain direction tensor\\	
		$\bm\Lambda^p_{fwd}$ \>\quad Reverse TRIP strain direction tensor\\

		$\bm{\varepsilon}^{t}$ \>\quad Transformation strain \\ 
		$\bm{\varepsilon}^{p}$ \>\quad TRIP strain \\ 		
		$\bm{\varepsilon}^{t-r}$ \>\quad Transformation strain at reverse point \\ 
		$\bm{\Upsilon}$ \>\quad Set of internal state variables\\
		$\bm{\sigma}$ \>\quad Cauchy stress \\ 
		$\bm{\sigma}_f$ \>\quad Effective stress \\ 	
		
		$\bm{\sigma}'$ \>\quad Deviatoric part of Cauchy stress \\ 
		$\bm{\sigma}_f'$ \>\quad Deviatoric part of effective stress \\ 		
		${\bar\sigma}$ \>\quad von Mises equivalent Cauchy stress \\  
		${\bar\sigma_f}$ \>\quad von Mises equivalent effective stress \\  		
		$\bm{\beta}$ \>\quad Internal stress \\ 		 
		$\bm\alpha$ \>\quad Effective thermal expansion tensor   \\
		$\Delta\alpha$ \>\quad Phase difference of thermal expansion \\      
		
		$A_s$ \>\quad Austenite transformation start temp.\\
		$A_f$ \>\quad Austenite transformation finish temp.\\  
		$C_1^p$ \>\quad TRIP strain evolution parameter\\  
		$C_2^p$ \>\quad TRIP strain evolution parameter\\  		
		$D$ \>\quad Smooth hardening function parameter\\  			
		$M_s$ \>\quad Martensite transformation start temp.\\
		$M_f$ \>\quad Martensite transformation finish temp.\\       
		$G$ \>\quad Gibbs free energy\\
		$H^{cur}$ \>\quad Current transformation strain\\  		
		$H^{max}$ \>\quad Maximum transformation strain\\  
		$T$ \>\quad Temperature \\ 
		$T_0$ \>\quad Temperature at reference point \\ 
		$Y$ \>\quad Critical thermodynamic driving force \\   
		$Y_0$ \>\quad Reference thermodynamic driving force \\ 		
		
		$a_1,a_2,a_3$ \>\quad Material parameters in hardening function \\ 
		$c$ \>\quad Specific heat \\       
		$f(\xi)$ \>\quad Smooth hardening function \\
		$n_1,n_2$ \>\quad Smooth hardening parameters \\	
		$n_3,n_4$ \>\quad Smooth hardening parameters \\				
		$k_t$ \>\quad Material parameter in $H^{cur}$ curve \\    		     
		$s$ \>\quad Effective specific entropy \\
		$s_0$ \>\quad Specific entropy at reference state \\  
		$\Delta s_0$ \>\quad Difference of specific entropy \\    
		$u$ \>\quad Effective internal energy\\
		$u_0$ \>\quad Internal energy at reference state\\  
		$\Delta u_0$ \>\quad Phase difference of internal energy \\

		$\Phi$ \>\quad Transformation function\\
		${\sigma}_b$ \>\quad Maximum internal stress magnitude \\ 			
		$\rho$ \>\quad Density \\
		$\xi$ \>\quad Martensitic volume fraction  \\
		$\xi^d$ \>\quad Detwnnied martensitic volume fraction  \\
		$\pi$ \>\quad Thermodynamic driving force\\  
		$\lambda_1$ \>\quad Internal stress evolution parameter\\  		
		$\zeta^d$ \>\quad Accumulated detwinned martensitic volume fraction\\

	\end{tabbing}
\end{multicols}

\section{INTRODUCTION}\label{sec:intro}

SMAs are metallic alloys with the ability to recover their pre-defined shapes when subjected to appropriate thermo-mechanical loadings \cite{lagoudas2008}. Since the discovery of their shape memory effect, SMAs have been intensively investigated as actuators to realize actively controlled morphing structures during the past several decades. 
Due to their high actuation energy density compared to other active materials, such as shape memory polymers and piezoelectrics, the current and potential applications of SMA-based actuators
in the biomedical, aerospace, automobile and energy fields are expanding rapidly. For instance, an SMA-based beam component was used as a bending actuator to morph the engine outer shell geometry so that desired aerodynamic conditions can be achieved for the airplane\cite{hartl2007aerospace}. Another example is a SMA-based torque tube that was used as a rotation actuator to deploy and contract a solar panel for small satellite\cite{wheeler2015}. 
SMA torque tubes are also used as rotational actuator installed on a Boeing airplane to morph the wing flaps at the trailing edge in a full scale flight test\cite{mabe2014,calkins2016}.

Experimental evidences indicate that the response of SMAs subjected to cyclic thermomechanical loading conditions is non-stable and TRIP strains are accumulated during the process\cite{Bo1999_TRIP_2,Bo1999_TRIP_3,lagoudas2004TRIP}. The reason is repeated phase transformation induces significant distortion located at the austenite-martensite interfaces and grain boundaries, especially during the very first thermomechanical cycles. This distortion drives dislocation activity resulting in an observable macroscopic transformation-induced plastic strain, which occurs at effective stress levels much lower than the plastic yield limit of the material. In general, more than 20\% TRIP strain may accumulate during the lifetime of an SMA component\cite{wheeler2014}. As a result, a training procedure, i.e., thermal cycling under external bias load, is commonly performed on untrained SMAs components to stabilize their response before they are used as actuators. 

A substantial number of consitutive models have been developed to predict the stable response of trained SMAs. A thorough literature review of the available work can be found from literatures \cite{boyd1996,lagoudas2008,lagoudas2012,Xu2017,arghavani2011,tham2001,patoor1996,patoor2006}. 
In the present work, a new three-dimensional SMA phenomenological constitutive model is developed to capture the evolutionary response of untrained SMAs. The proposed model predicts the traditional thermomechanically induced phase transformations, and accounts for the evolutionary material response due to the accumulation of TRIP strains. This model is also able to predict the generation of internal stresses due to the presence of the TRIP strain, which in return leads to its capability to predict the TWSME exhibited by trained SMAs under stress-free thermal cycling loading conditions. The martensitic volume fraction, transformation strain tensor, TRIP strain tensor, and internal stress tensor are used as internal state variables in the proposed modeling framework, and the predictions provided by the model are compared with the corresponding experimental results to demonstrate its capabilities. The proposed model is anticipated to be further validated against additional experimental data for NiTiHf SMAs under general complex loading conditions. The validated model will be utilized as an efficient tool for the design and analysis of SMA-based actuators, such as SMA beams and torque tubes, which are intended to be integrated with the future supersonic transport aircraft   that can adapt their shape to reduce the sonic boom noise from takeoff to landing.  


\vspace{0.5cm}
\begin{figure}[h!]	
	\begin{center}
		\includegraphics[width=0.65\columnwidth]{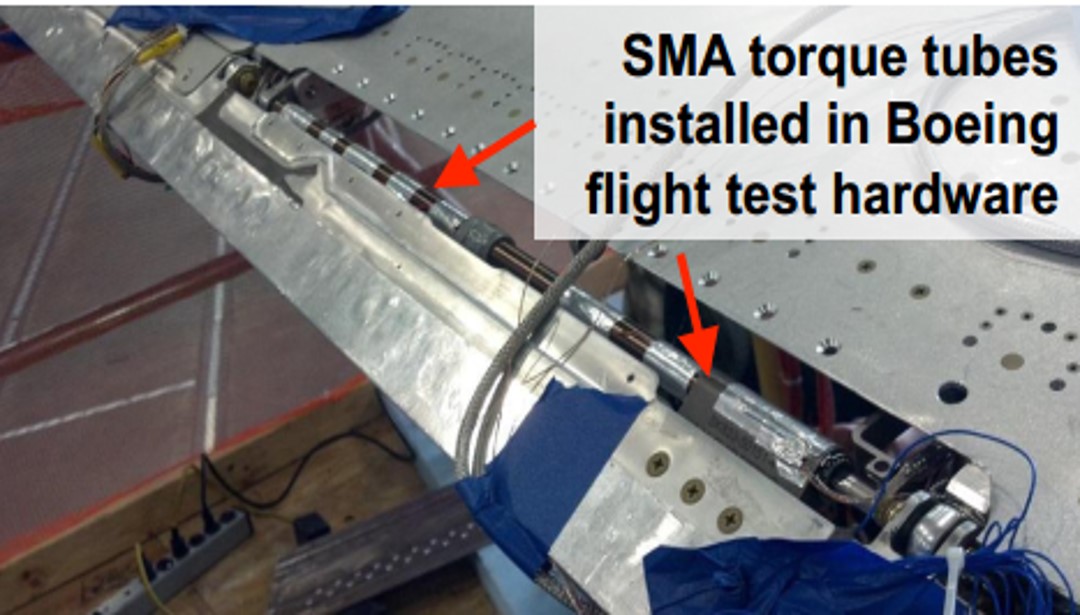}		
	\end{center}
	\caption{An SMA torque tube actuator installed on a Boeing airplane to rotate the trailing edge wing flaps for a full scale flight test\cite{mabe2014,calkins2016}.}
	\label{fig:Boeing_wing}
\end{figure}
  
\section{Model Formulation}\label{Model}
\subsection{Constitutive model for SMAs}
\subsubsection{Thermodynamic potential and constitutive equations}
Based on the work of Lagoudas et al.\cite{lagoudas2004TRIP} and Xu et al.\cite{Xu2017trip}, the Gibbs free energy $G$ is proposed to be a continuous function dependent on Cauchy stress tensor $ \bm{\sigma} $, temperature $ T $, and a set of internal state variables $\mathbf{\Upsilon}=\{\bm{\varepsilon}^{t},\bm{\varepsilon}^{p},\bm{\beta},\xi\}$, i.e., the transformation strain tensor $ \bm{\varepsilon}^{t} $, the TRIP strain tensor $ \bm{\varepsilon}^{tp} $, the internal stress tensor $\bm{\beta}$, and the martensitic volume fraction scalar $ \xi $, respectively. The $ \mathbf{\varepsilon}^{t} $ is used to account for the inelastic strain caused by the phase transformation, $\mathbf{\varepsilon}^{p}$ is used to represent the large accumulated plastic strain induced by phase transformation, $\bm{\beta}$ is used for the generated internal stress tensor during training process, and the martensitic volume fraction $ \xi $ (ranging $ 0 \leqslant \xi \leqslant 1 $) is used for differentiating the two different phases in the SMAs system. The explicit expression for $G$ is given as follows,
\begin{equation}\label{GIBBS_explicit} 
\begin{aligned}
G =  -\dfrac{1}{2 \rho} \bm{\sigma} : \mathcal{S}\bm{\sigma} - \dfrac{1}{\rho}  \bm{\sigma} :[~\bm{\alpha}(T-T_0)+\bm{\varepsilon}^{t}+\bm{\varepsilon}^{p}]- \dfrac{1}{\rho}\int^{\xi}_0 (\bm{\beta}:\frac{\partial\bm{\varepsilon}^{t}}{\partial\xi} )d\xi  
+c \Big[(T-T_0)-T\ln (\dfrac{T}{T_0}) \Big]-\\s_0 T+u_0+f(\xi)
\end{aligned}
\end{equation}
in which, $ \mathcal{S}$ is the effective fourth-order compliance tensor calculated by equation (\ref{eq:S_mix}), $\bm{\alpha}$ is the effective second order thermal expansion tensor, $ c $ is effective specific heat, $ s_0, u_0 $ are effective specific entropy and effective specific internal energy at reference state, respectively, $ T $ is the current temperature and $ T_0 $ is the reference temperature, $f(\xi)$ is a smooth hardening function.
\begin{equation}\label{eq:S_mix}
\mathcal{S}(\xi)=\mathcal{S}^A + \xi(\mathcal{S}^M-\mathcal{S}^A)=\mathcal{S}^A + \xi\Delta\mathcal{S}
\end{equation}

On basis of the proposed Gibbs free energy, following classical thermodynamic principles and standard Coleman-Noll procedure, the following constitutive relations can be obtained,
\begin{equation}\label{eq:Cons_stress}
\bm \varepsilon =  -\rho\frac{\partial G}{\partial \bm {\sigma}} = \mathcal{S}\bm{\sigma} + \bm\alpha(T-T_0)+ \bm {\varepsilon}^{t}+ \bm{\varepsilon}^{p}
\end{equation}   
\begin{equation}\label{eq:Cons_entropy}
s =  -\rho_{0}\frac{\partial G}{\partial T} = \frac{1}{\rho}  \bm{\sigma} : \bm\alpha  + c \ln (\dfrac{T}{T_0}) -s_0
\end{equation}
%
\subsubsection{Evolution equation for transformation strain}\label{Evolution_Trans}
In this part, the evolution equation of transformation strain $\bm{\varepsilon}^t$ is presented. Following the principle of maximum dissipation\cite{qidwai2000}, it is proposed that the rate change of the transformation strain ${\dot{\bm \varepsilon}}^{t}$ is proportional to the rate change of the martenstitic volume fraction $\dot{\xi}$, which results in the following evolution relationship between $\xi$ and  $ \bm{\varepsilon}^t $,
\begin{equation}\label{eq:Trans_Evol}
\begin{aligned}
{\dot{\bm \varepsilon}}^{t}= \bm{\Lambda}  \dot{\xi},  \ \  \bm\Lambda=\begin{cases}\bm{\Lambda}_{fwd}, \; \dot{\xi}>0, \vspace{5pt} \\ \bm{\Lambda}_{rev}, \; \dot{\xi}<0, \end{cases}\\
\end{aligned}
\end{equation}
where $\bm\Lambda_{fwd}$ is called the forward transformation direction tensor and $\bm\Lambda_{rev}$ is the reverse transformation direction tensor, they are defined as follows respectively, 
\begin{equation}\label{eq:Evo_tr}
\bm\Lambda_{fwd}=\frac{3}{2} H^{cur} \dfrac {{ \bm{\sigma}_f^{'}} } { {{\bar{\bm{\sigma}}}_f} }   \ \     
\bm\Lambda_{rev}=\dfrac{\bm\varepsilon^{t-r}}{\xi^r} .
\end{equation}
in which $ \bm{\sigma}_{{f}} $ is called the effective stress tensor defined in equation (\ref{eq:stress_eff}), $\bm\varepsilon^{\textit{t-r}}$ and ${\xi}^{r}$ denotes the value of transformation strain and martensitic volume fraction at the reverse transformation starting point. The internal stress tensor $\bm\beta$ generated during cyclic loading is introduced in section \ref{sec:Evo_B_stress}. 
\begin{equation}\label{eq:stress_eff}
\bm{\sigma}_{f}=\bm{\sigma}+\bm\beta
\end{equation}
$H^{cur}$ denotes the current maximum transformation strain at current stress state. Based on the observation from experimental results, the transformation strain is usually not a constant value but exponentially depends on current material stress state. The reason is that the austenitic phase can be transformed into oriented/twinned martensitic phase depending on the current stress state. Therefore an exponential function $H^{\textit{cur}}$ based on current stress state is introduced to obtain the current transformation strain magnitude as shown in equation (\ref{eq:Hcur}), where $H^{\textit{max}}$ is the maximum (or saturated) transformation strain and $\textit{k}_t$ is a material curve fitting parameter.
\begin{equation}\label{eq:Hcur}
H^{cur}(\bm\sigma)= H^{max}(1-e^{-{k}_t { \bar{\bm\sigma}_{f}}})
\end{equation}

$ \bm{\sigma}^{'}_{f} $ is the deviatoric part of the effective stress tensor, i.e., {\small $ \bm{\sigma}^{'}_{f} =\bm{\sigma}_{f} -{\small \frac{1}{3}}\textrm{tr}(\bm{\sigma}_{f})*\mathbf{1} $}, in which $ \mathbf{1} $ is the second order identity tensor. The von mises equivalent effective stress $\bar{\sigma}_{f} $ is given as,
\begin{equation}
\bar{\sigma}_{f} ={\small \sqrt{{{\small \frac{3}{2}}\bm\sigma^{'}_{f}}:\bm{\sigma}^{'}_{f}}} 
\end{equation}

\subsubsection{Evolution equation for TRIP strain}\label{Evolution_Plastic}
In this part, the evolution equation for the TRIP strain ${\bm \varepsilon}^{p}$ is presented. Based on the observation from experimental results, the TRIP strain is evolved with respect to the loading cycles. Following the evolution equation proposed by Lagoudas and Entchev \cite{lagoudas2004TRIP}, it is proposed that the rate change of the TRIP strain ${\dot{\bm \varepsilon}}^{p}$ is proportional to the rate change of the martenstitic volume fraction $\dot{\xi}$, which is expressed as follows,
\begin{equation}\label{eq:Evol_plastic}
\begin{aligned}
{\dot{\bm \varepsilon}}^{p}= \bm{\Lambda}^p  \dot{\xi},  \ \  \bm\Lambda^p=\begin{cases}\bm{\Lambda}^p_{fwd}, \; \dot{\xi}>0, \vspace{5pt} \\ \bm{\Lambda}^p_{rev}, \; \dot{\xi}<0, \end{cases}\\
\end{aligned}
\end{equation}
in which $\bm\Lambda^p_{fwd}$ is called the forward TRIP direction tensor,  $\bm\Lambda^p_{rev}$ is the reverse TRIP direction tensor. Their explicit expressions are defined as follows, 

\begin{equation}\label{Dir_plastic}
\bm\Lambda^p_{fwd}=\frac{3}{2} C^p_1 \dfrac{H^{cur}(\bm\sigma)}{H^{max}}\dfrac{{ \bm{\sigma}_f^{'}}} {{{\bar{\bm{\sigma}}}_f} } e^{-\frac{\zeta^d}{C_2^p}}  ;\ \ 
\bm\Lambda^p_{rev}=- C^p_1 \dfrac{H^{cur}(\bm\sigma)}{H^{max}}\dfrac{\bm \varepsilon^{t-r}}{\xi^r} e^{-\frac{\zeta^d}{C_2^p}}.
\end{equation}

The evolution of TRIP strain assumes that only the orientated (or called detwinned) martensitic phase transformation contributes to the generation of TRIP stain while the twinned martensitic phase transformation does not \cite{lagoudas2004TRIP}. This assumption is reflected in the above TRIP direction tensor by the ratio of $H^{cur}$ over ${H^{max}}$ and a term $\zeta^d$ called the accumulated detwinned martensitic volume fraction in the exponential function. The explicit expression of $\zeta^d$ is defined as follows,
\begin{equation}
\zeta^d = \int_0^t |\dot{\xi}^d(t)|dt
\end{equation}
in which the detwinned martensitic volume fraction $\xi^d$ is obtained by,
\begin{equation}\label{eq:xid}
\xi^d = \dfrac{H^{cur}(\bm\sigma)}{H^{max}}\xi
\end{equation}
%

\subsubsection{Evolution equation for internal Stress}\label{sec:Evo_B_stress}
As described in section \ref{sec:intro}, the mechanism for promoting TWSME in SMAs is the generation of internal stresses by inducing the microstructure changes (such as dislocation movement, retained martensitic variants, and oriented growth of coherent precipitates) through cyclic loading\cite{atli2015}. These generated oriented internal stress fields along with a potential thermal stimulus are then able to drive the oriented martensitic phase transformation when SMAs subject to stress free conditions resulting in the so-called TWSME. In oder to model this process, a second order internal stress tensor $\bm{\beta}$ is introduced in this model. The evolution equation of this internal stress is proposed as follows,
\begin{equation}\label{eq:internal_stress}
\bm{\beta} =  {\sigma}_b~ \dfrac{\bm \sigma_f}{\bar{\bm \sigma }_f} (1-e^{-\lambda_1\zeta^d}
) 
\end{equation}
this evolution equation assumes that the internal stress is generated along the same direction of the applied external forces, in which $\sigma_b$ is a material parameter representing the maximum (or saturated) internal stress magnitude that can be generated during the training process. $\zeta^d$ is the accumulated detwinned martensitic volume fraction described above and $\lambda_1$ is a curve fitting parameter indicating how the internal stress evolves with respect to the number of loading cycle.
\subsubsection{Smooth hardening function}\label{Smooth_function}
A smooth hardening function $f(\xi)$ is proposed in equation (\ref{eq:Smooth_hardeing}) to account for the hardening effects associated with the transformation process \cite{lagoudas2008}, where three additional intermediate material parameters $a_1,a_2,a_3$ and four curve fitting parameters $n_1,n_2,n_3,n_4$ are introduced to better treat the smooth transition behaviors during the initiation and completion of  transformations.
\begin{equation}\label{eq:Smooth_hardeing}
\begin{aligned}
f(\xi)=   \begin{cases} \cfrac{1}{2} a_1\Big(  \xi  + \frac{\xi^{n_1+1}} {n_1+1}+ \frac{(1-\xi)^{n_2+1}} {n_2+1} \Big)+a_3\xi ~, \; \dot{\xi}>0, \vspace{5pt} \\ 
\cfrac{1}{2} a_2\Big(  \xi  + \frac{\xi^{n_3+1}} {n_3+1}+ \frac{(1-\xi)^{n_4+1}} {n_4+1} \Big)-a_3\xi ~, \; \dot{\xi}<0 \end{cases}\\
\end{aligned} 
\end{equation}
%


\subsubsection{Transformation function}\label{Trans_Func}
The objective of this section is to define a proper transformation criterion to determine the occurrence of the phase transformations. After the introduction of evolution equation for internal state variables, on basis of the $2^{nd}$ thermodynamic principle, substituting the evolution equation (\ref{eq:Trans_Evol}) and (\ref{eq:Evol_plastic}) into the reduced form dissipation inequality resulting in the following expression,
\begin{equation}\label{Dissipation_xi}
\begin{aligned}
\big(\bm\sigma:\bm\Lambda  + \bm\beta:\bm\Lambda + \bm\sigma:\bm\Lambda^p- 
\rho\frac{\partial G}{\partial \xi}\big)\dot{\xi}=\pi\dot{\xi}\geqslant 0 
\end{aligned}
\end{equation}
where the quantity $\pi$ is called the general thermodynamic driving force conjugated to $\xi$. Substituting the equation (\ref{GIBBS_explicit}) of Gibbs free energy into equation (\ref{Dissipation_xi}), the explicit expression for the thermodynamic driving force $ \pi $ is obtained as,
\begin{equation}\label{eq:pi}
\begin{aligned}
\pi=(\bm\sigma+ \bm\beta):\bm\Lambda + \bm\sigma:\bm\Lambda^p+
\dfrac{1}{2}\bm\sigma:{\Delta}\mathcal{S}\bm\sigma+\bm\sigma:{\Delta}\bm{\alpha}(T-T_0)+ \rho\Delta s_0 T -\rho\Delta c 
\big[ T-T_0-T\ln(\dfrac{T}{T_0}) \big ] - \\ \rho\Delta u_0 - \frac{\partial f}{\partial \xi}
\end{aligned}
\end{equation}
where $\Delta \mathcal{S},\Delta \bm{\alpha}, \Delta c, \Delta s_0$, and $ \Delta u_0 $ are the phase differences on compliance tensor, thermal expansion tensor, specific heat, reference entropy and reference internal energy. It can be observed that the thermodynamic driving force $\pi$ is a function of stress $\bm{\sigma}$, temperature $T$ and martenstic volume fraction $\xi$. This indicates that the phase transformation process can be activated by two independent sources, namely either the stress or temperature, which correlates quite well with the experimentally observed stress-induced and thermally-induced phase transformations in SMAs. To proceed to the goal of defining a transfomration criteria, it is assumed that whenever the thermodynamic driving force $\pi$ reaches a critical value $Y $ ($ -Y $), the forward (reverse) phase transformation takes place. Therefore a transformation function $\Phi$ can be defined as the transformation criteria to determine the transformation occurrence as follows,
\begin{equation}\label{eq:Trans_Fun}
\normalfont{\Phi}=\begin{cases}~~\pi - Y, \; \dot{\xi}>0, \vspace{5pt} \\ -\pi - Y, \; \dot{\xi}<0, \end{cases}\\
\end{equation}
In the continuous development of the established SMA model\cite{lagoudas2012}, a reference critical value $Y_0$ and an additional parameter D were introduced into $Y$ such that the model can capture the different slopes $C_A, C_M$ in the effective stress-temperature phase diagram. This development is also adopted here.
\begin{equation}\label{Critical_Y}
Y(\bm{\sigma}) = \begin{cases}Y_0 + D\bm\sigma:\bm\Lambda_{\textit{fwd}}, \; \dot{\xi}>0, \vspace{5pt} \\ Y_0 + D\bm\sigma:\bm\Lambda_{\textit{rev}}, \; \dot{\xi}<0, \end{cases}\\
\end{equation}
As a consequence of the maximum dissipation principle, the so-called Kuhn-Tucker constraints are applied on the proposed model as follows\cite{qidwai2000},
\begin{equation}\label{eq:Kuhn-Tucker}
\begin{aligned}
\dot{\xi} \geqslant 0; \quad \Phi(\bm\sigma,T,\xi)= ~~\pi - Y \leqslant 0;  \quad  \Phi\dot{\xi}=0;~~~\bf{(A\Rightarrow M)}\\
\dot{\xi} \leqslant 0; \quad \Phi(\bm\sigma,T,\xi)= -\pi - Y \leqslant 0; \quad   \Phi\dot{\xi}=0;~~~ \bf{(M\Rightarrow A)}
\end{aligned}
\end{equation}

\subsubsection{Determination of material parameters}
There are three groups of material parameters need be calibrated described as table \ref{tab:Material_bar}. Elastic modulus $ E_A, E_M $ can be determined through a pseudoelastic test by calculating the slopes at initial loading and unloading stage. A widely accepted assumption for Poisson's ratio is $\nu_A=\nu_M$ with reported value of $0.33$ found in literature \cite{Bo1999_TRIP_2}. It is also reasonable to assume that the thermal expansion coefficient for two phases are equal $\alpha_A=\alpha_M$ which can be calibrated through an isobaric actuation experiment. Other material constants characterizing stable transformation cycles $(C_A, C_M, M_s, M_f, A_s, A_f)$ can be calibrated through the stress-temperature phase diagram. Next, material parameters related to transformation strain and smooth hardening effects are discussed. The maximum transformation strain $H^{max}$ can be directly determined from experimental data and the value of parameter $k_t$ are chosen to best fit the $H^{cur}$ curve. Coefficients $n_1, n_2, n_3, n_4$ without specific physical material meanings are determined to best match the smoothness corners of transformation hysteresis. Material parameters related to TRIP strain and internal stress are determined from experiment data. There are also seven additional parameters $(\rho\Delta s_0, \rho\Delta u_0, a_1, a_2,a_3, Y_0 ~\text{and} ~D)$ that can be written by the aforementioned material parameters. Referring to the literature\cite{lagoudas2012}, the first 5 intermediate material parameters $(a_1, a_2, a_3, \rho\Delta u_0,  Y_0)$ can be expressed in terms of the previous directly measured material parameters as,
\begin{equation}\label{eq:Reduced_5_Parameters}
\begin{aligned}
\begin{cases}
&a_1=\rho \Delta s_0 (M_f-M_s); \quad a_2=\rho \Delta s_0 (A_s-A_f)\\[6pt]
&a_3 = \cfrac{1}{4}~a_2(1+\cfrac{1}{n_3+1})-\cfrac{1}{4}~a_1(1+\cfrac{1}{n_1+1})\\[6pt]
&\rho \Delta u_0 =\dfrac{1}{2}\rho \Delta s_0 (M_s+A_f) + \beta\Lambda\\[6pt]
&Y_0=\dfrac{1}{2} \rho \Delta s_0 (M_s-A_f)-a_3\\[6pt]
\end{cases}
\end{aligned}
\end{equation}
the additional two intermediate material parameters $\rho\Delta s_0$ and $D$ can be expressed as, \vspace{0.3cm}
%
\begin{equation}\label{eq:Reduced_D}
\begin{aligned}
\begin{cases}
& D = \dfrac{ (C_M-C_A)\big[ H^\textit{cur}+ \sigma\partial_{\sigma}H^\textit{cur}+\beta\partial_{\sigma}H^\textit{cur}  + \sigma(\frac{1}{E_\textit{M}}-\frac{1}{E_ \textit A })\big]}
{(C_M+C_A) ( H^\textit{cur} + \sigma\partial_{\sigma}H^\textit{cur})}
+\dfrac{C_1^p }{H^\textit{max}} e^{-\frac{\zeta^d}{C_2^p}} \\[14pt]
& \rho \Delta s_0 = -\dfrac{ -2C_M C_A \big[ H^\textit{cur}+ \sigma\partial_{\sigma}H^\textit{cur}+\beta\partial_{\sigma}H^\textit{cur}  + \sigma(\frac{1}{E_\textit{M}}-\frac{1}{E_ \textit A })\big]}
{C_M+C_A} 
\end{cases}
\end{aligned}
\end{equation}  

\vspace{0.3cm}

\section{NUMERICAL RESULTS}\label{Result}
The model has been proposed in the previous sections and implemented though an user-defined material subroutine(UMAT) under the commercial finite element software Abaqus. To demonstrate the capabilities of the proposed model to capture the TRIP strains and internal stresses generated during thermomechanical cyclic loading, two boundary value problems (BVPs) are investigated here and the model predictions are compared against the available experimental data. The first problem analyzes the response of a NiTi SMA under isothermal loading conditions subjected to 50 mechanical cycles. In this BVP, the stress-induced phase transformation is simulated and the pseudoelastic stress-strain curves are obtained. The second BVP analyzes the respone of a NiTi SMA under isobaric loading condition subjected to 100 thermal cycles. The thermally-induced phase transformation is simulated and the actuation strain-temperature curves are predicted. The simulations are performed by using the commercial finite element software Abaqus 6.14\cite{abaqus2014}.

\subsection{Isothermal Loading Case}\label{sec:uniaxial_isothermal}
\subsubsection{SMAs under tensile loading}\label{sec:Strip_isothermal}
The first BVP considers an untrained NiTI SMA loaded under isothermal loading conditions and subjected to 50 mechanical cycles. During the mechanical loading cycles the material is loaded up to 550 MPa and then is unloaded to 0 MPa. The temperature kept at a constant at $360$ K throughout the entire process. During this process, the material undergoes a forward stress-induced phase transformation during the mechanical loading phase followed by a reverse phase transformation during the unloading phase. The material parameters used in this simulation are summarized in table \ref{tab:Material_bar}. 

\begin{figure}[tb!]
	\begin{center}
		\advance\leftskip-0.5cm
		\includegraphics[width=0.9\columnwidth]{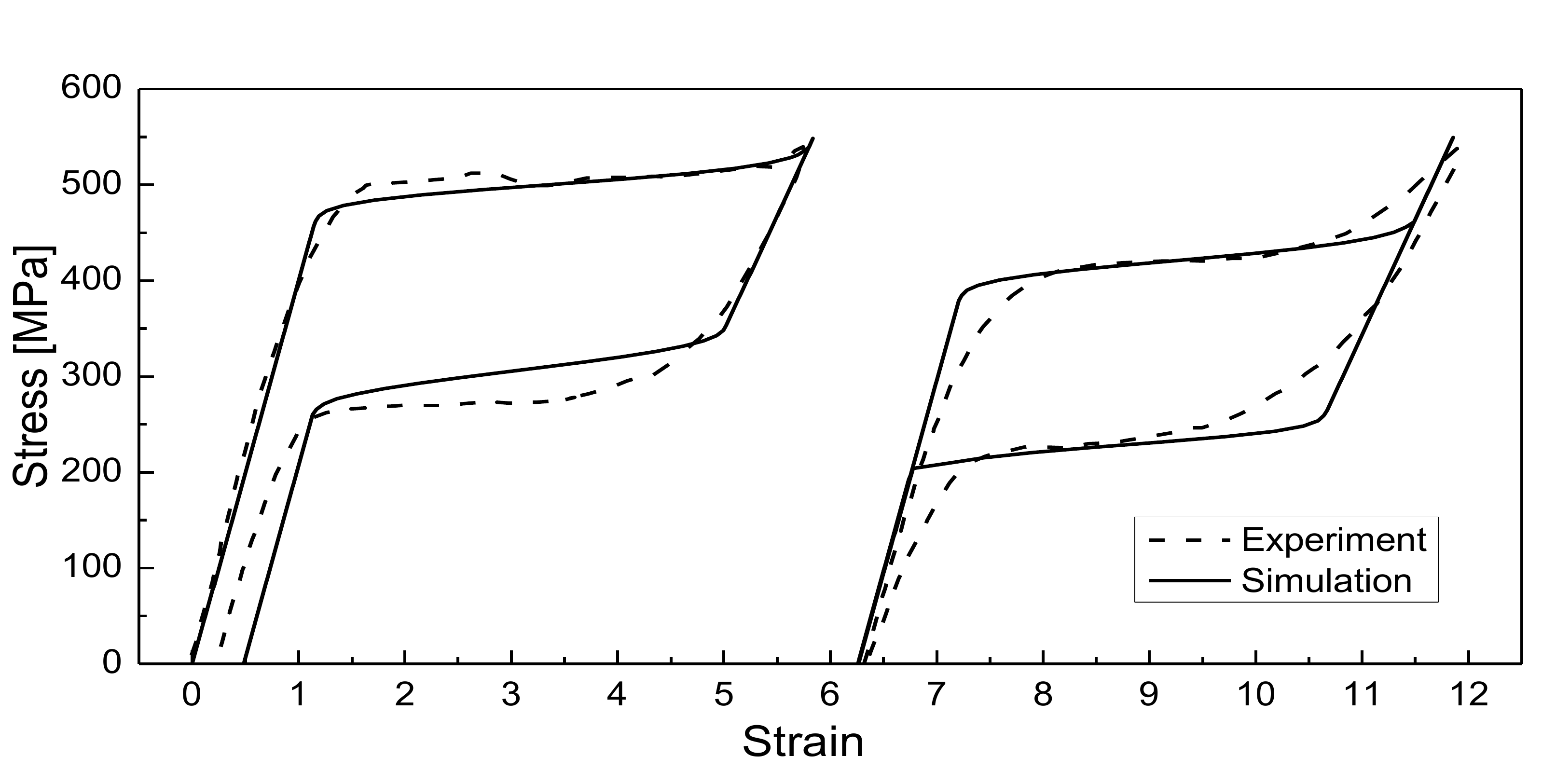}
	\end{center}
	\vspace{-0.5cm}
	\caption{The $1^{st}$ and $50^{th}$ pseudoelastic stress-strain response of the untrained NiTi under 550 MPa uniaxial tensile loading.}
	\label{fig:isothermal}	
\end{figure}

\begin{figure}[tb!]
	\begin{center}
		\includegraphics[width=0.6\columnwidth]{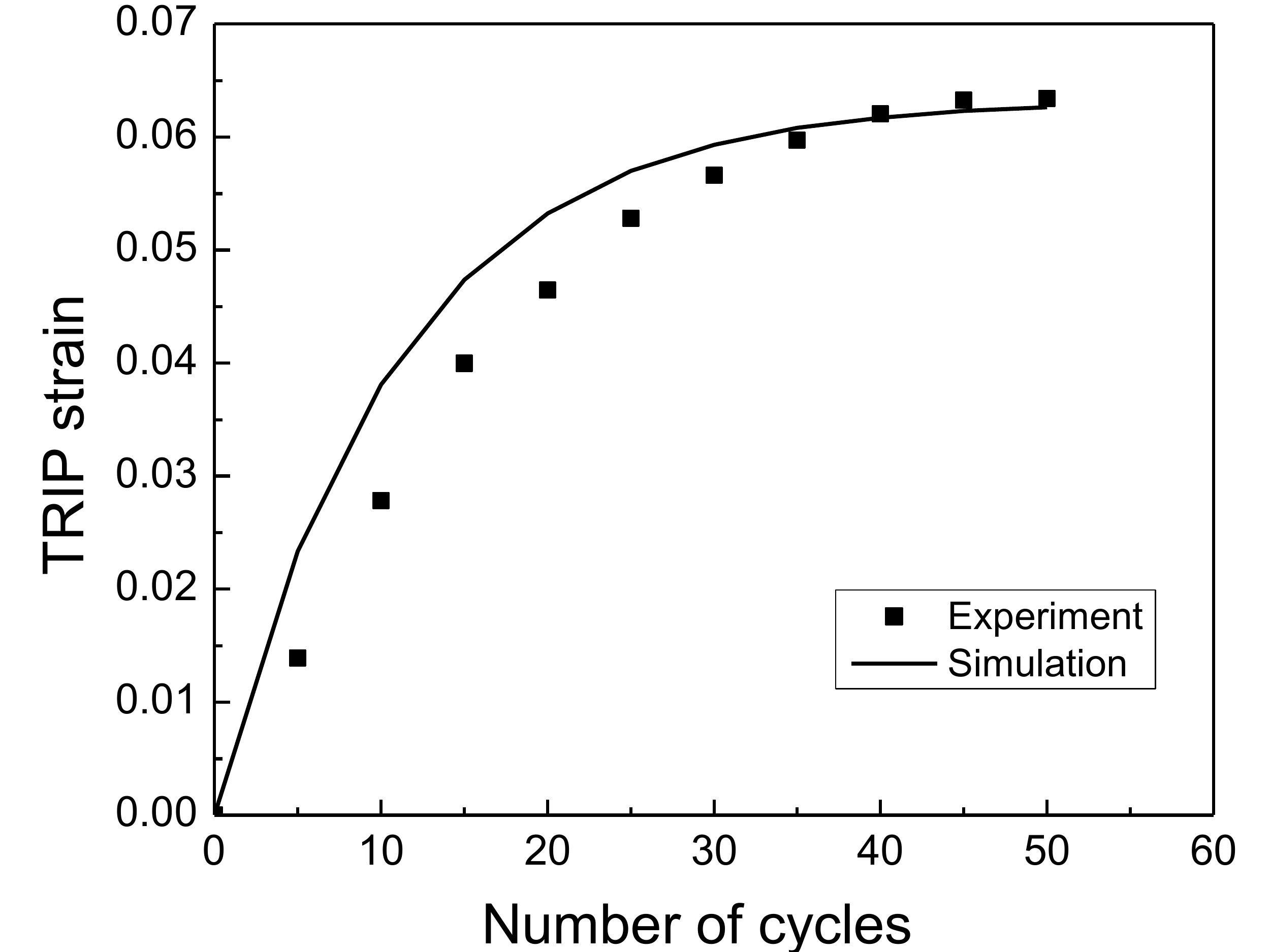}
	\end{center}
	\vspace{-0.3cm}
	\caption{The evolution of TRIP strain for the untrained NiTi material subject to 50 isothermal loading cycles.}	\label{fig:TRIP_number}
\end{figure}

The stress-strain curves obtained by the propsed model are compared with the available experimental data \cite{lagoudas2008}. As shown in figure \ref{fig:isothermal}, the NiTi SMA accumulated a large amount of irrecoverable TRIP strain from the $1^{st}$ cycle to the $50^{th}$ cycle and this behavior is well captured by the proposed model.  Moreover, it is interesting to note that the model was able to capture the experimentally observed decrease of the stress level at which the forward phase transformation starts as a result of the mechanical cycling. This intrinsic behavior predicted by the proposed model using
the internal stress term in the formulation of the model.The TRIP strain accumulated with respect to number of loading cycles is also plotted in figure \ref{fig:TRIP_number}. It is shown that the TRIP strain generated drastically within the initial 30 loading cycles but has stabilized afterwards. The TRIP strain in the $1^{st}$ cycle is around 0.7\% and  with a saturated value around 6.5\% after 30 cycles. The predicted TRIP evolution result through the proposed model is in good agreement with the experimental data.

\begin{table}[tb!] 
	\caption{Material parameters used for the NiTi SMA subjected to isothermal loading conditions.\cite{lagoudas2008}}
	\label{tab:Material_bar}\vspace{-0.5cm}
	\renewcommand{\arraystretch}{0.8}
	\begin{center}
		\begin{tabular}{c|lr|lr} \toprule
			Type                         &Parameter                        & Value                                   &Parameter            & Value  \\                                       \midrule
			&$E_A$                            & 70   [GPa]                              & $C_A$                & 3.4  [MPa/K]\\
			&$E_M$                            & 50   [GPa]                              & $C_M$               & 3.4  [MPa/K]\\
			Major material parameters      &$\nu_A=\nu_M$                          & 0.33                                    & $M_s$                & 264  [K]\\
			10                      		&$\alpha_A=\alpha_M$    &    2.2$\times$10$^{-5}$ [K$^{-1}$]                           & $M_f$                &160  [K]\\
			& $ H^\textit{max}$             & 0.05     & $A_s$                 &217  [K]\\
			&   $k_t$                		&    0.00752                  & $A_f $                 & 290  [K]\\                                   \midrule
			
			Smooth hardening parameters              	& $n_1$              & 0.2  & $n_3$         &  0.4\\
			6 & $n_2$         & 0.3 & $n_4$     &  0.5   \\  \midrule
			
			&$\sigma_b$                        & 100    [MPa]                      & $\lambda_1$                & 3.5 \\
			TRIP and Internal stress parameters               &$C_1^p$                      & 3.6$\times$ 10 $^{-3}$                      &                  &   \\
			4                     &$C_2^p$                      & 18                                 &                 &  \\                              
			\bottomrule
		\end{tabular}
	\end{center}
\end{table}

	\vspace{-0.5cm}

\begin{table}[t!] 
	\caption{Material parameters used for the untrained NiTi SMA subject to isobaric loading calibrated based on the experimental data from Atli et al.\cite{atli2015}.}
	\label{tab:Material_bar_2}\vspace{-0.5cm}
	\renewcommand{\arraystretch}{0.8}
	\begin{center}
		\begin{tabular}{c|lr|lr} \toprule
			Type                         &Parameter                        & Value                                   &Parameter            & Value  \\                                       \midrule
			&$E_A$                            & 22   [GPa]                              & $C_A$                & 16  [MPa/K]\\
			&$E_M$                            & 22   [GPa]                              & $C_M$                & 8   [MPa/K]\\
			Major material parameters       &$\nu_A=\nu_M$                          & 0.33                                    & $M_s$                & 340  [K]\\
			10                      		&$\alpha_A=\alpha_M$    &    2.2$\times$10$^{-5}$ [K$^{-1}$]                           & $M_f$                &310  [K]\\
			& $ H^\textit{max}$    	&  0.045    & $A_s$                 &361  [K]\\
			& $k_t$                 & 0.0158          & $A_f $                 & 395  [K]\\                                   \midrule
			
			Smooth hardening parameters 			& $n_1$          &     0.5              &   $n_3$        & 0.5 \\
			6								& $n_2$          &     0.5              &   $n_4$        & 0.5 \\  \midrule
			
			&$\sigma_b$                       & 50    [MPa]                      & $\lambda_1$                & 0.1 \\
			TRIP and Internal stress parameters         &$C_1^p$                      & 4.24$\times$ 10 $^{-3}$                      &                  &   \\
			4                             &$C_2^p$                      & 21.2                                 &                 &  \\                              
			\bottomrule
		\end{tabular}
	\end{center}
\end{table}

\vspace{0.5cm}
\subsection{Isobaric Loading Case}\label{uniaxial_isobaric}
The second BVP considers an untrained SMA loaded under isobaric loading conditions and subjected to 100 thermal cycles. To this end, the SMA is initially loaded under a constant tensile load of 200 MPa and then undergoes thermal cycles where the temperature varies from $30$ $\degree $C to $165$ $\degree $C for 100 cycles. During the aforementioned loading procedure the SMA undergoes a forward thermally-induced phase transformation followed by a reverse phase transformation. Once the 100 thermal cycles are completed, 10 additional thermal cycles under stress-free conditions are performed to examine the TWSME of the material. The material parameters used in this simulation are summarized in table \ref{tab:Material_bar_2}, while further details about the experimental data used are described in Atli et al.\cite{atli2015}.

\begin{figure}[H]	
	\begin{center}
		\includegraphics[width=0.7\columnwidth]{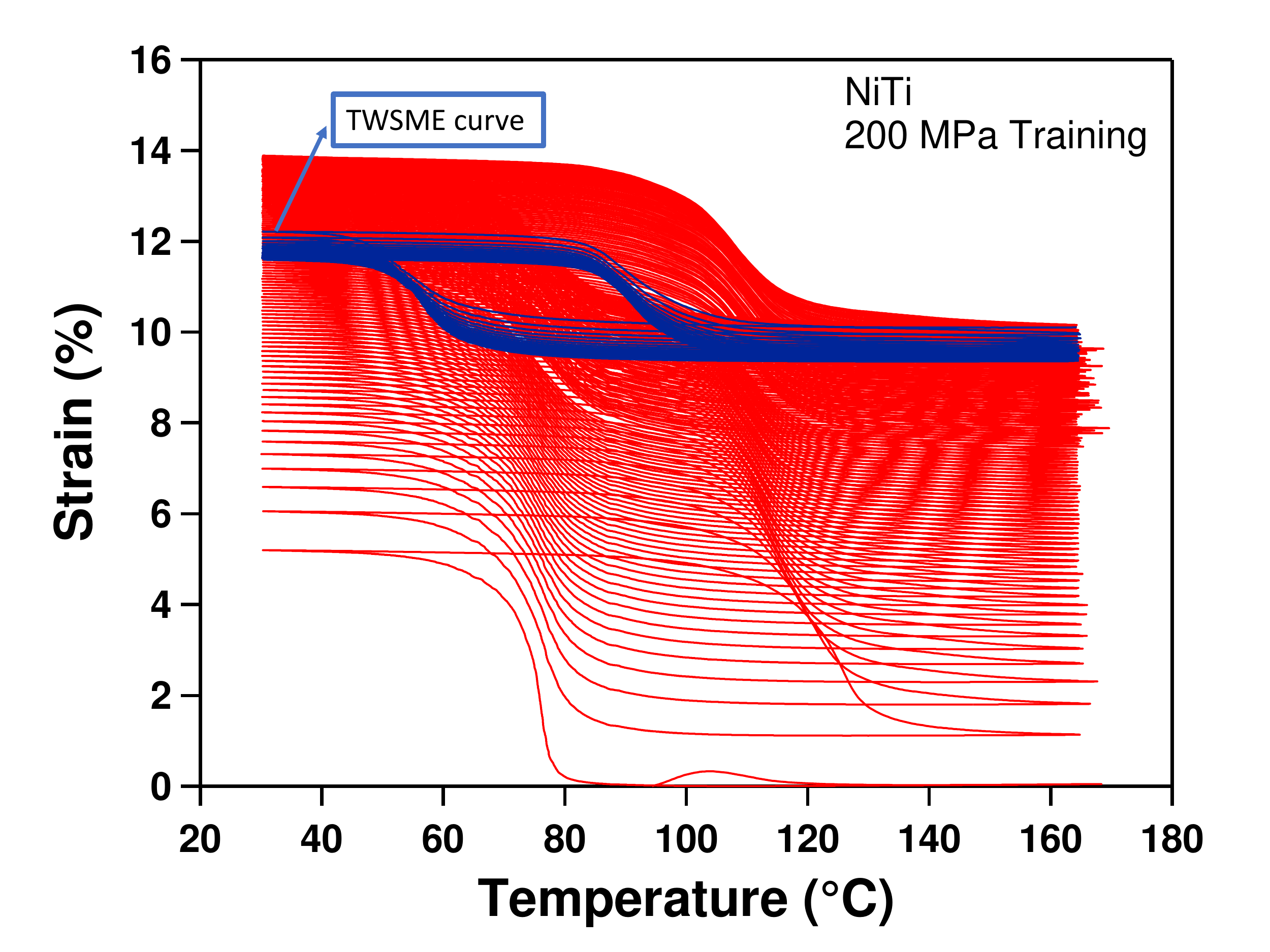}		
	\end{center}
	\vspace*{-0.5cm}
	\caption{Experimental result shows the TWSME at stress-free conditions for untrained NiTi material after 100 cycles of isobaric loading. The red curve is the response under bias load, and the blude curve indicates the reponse under stress-free condition.}
	\label{fig:TWSME_EXP}
\end{figure}

\begin{figure}[H]	
	\begin{center}
		\includegraphics[width=0.7\columnwidth]{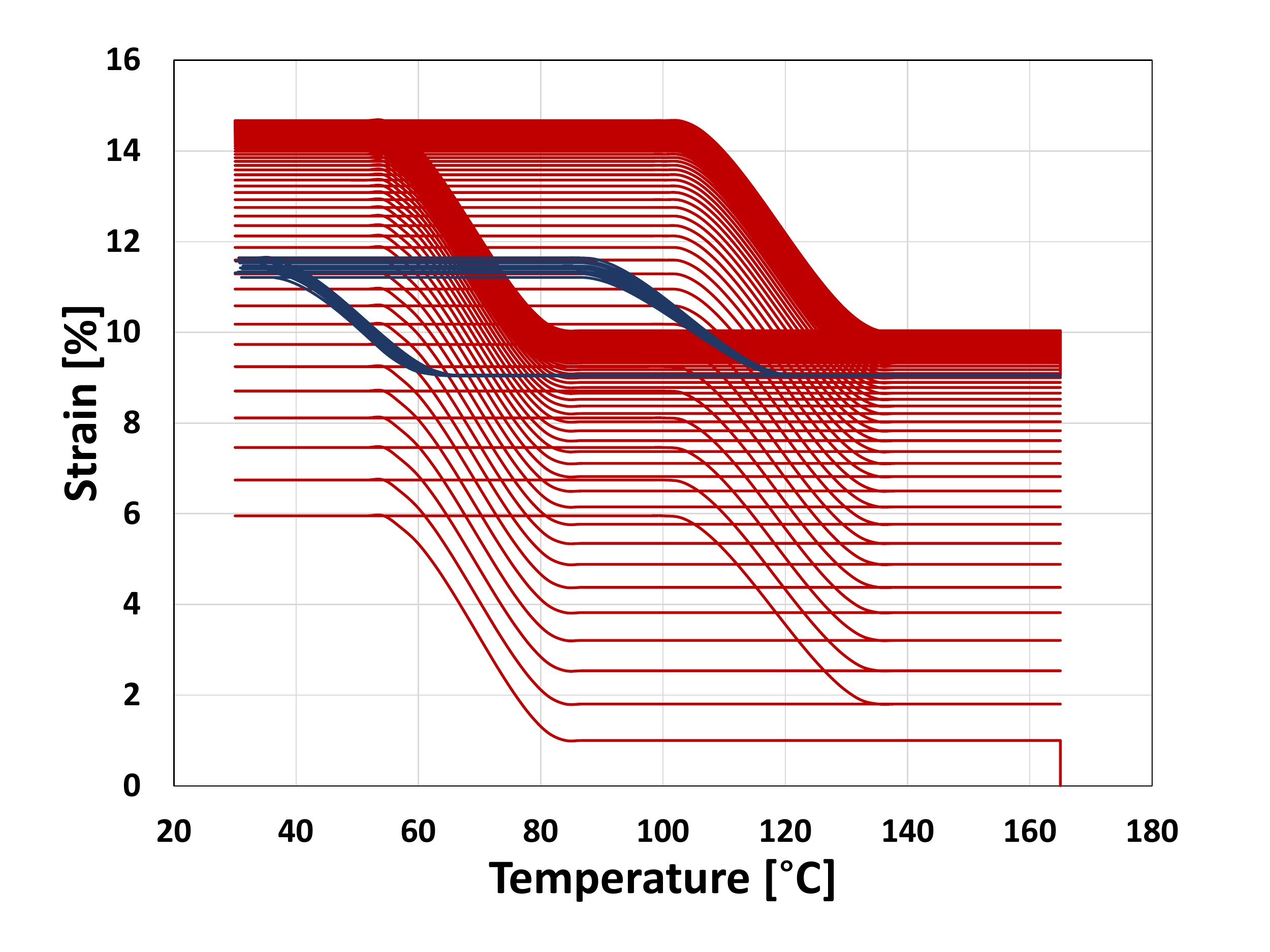}
	\end{center}
	\vspace*{-0.7cm}
	\caption{The cyclic temperature-strain response predicted by the proposed model for the untrained NiTi material subject to 100 isobaric loading cycles followed by 10 stress free loading cycles. The red curve is the response under bias load, and the blude curve indicates the reponse under stress-free condition.}
	\label{fig:TWSME_Simulation}
\end{figure}

The experimental result is shown in figure \ref{fig:TWSME_EXP}, in which the red curve represents the response of the SMA under bias load while the blue curve indicates its response under stress-free condition. It can be seen that a large amount of TRIP stain (around $10\%$) is accumulated during the 100 thermal loading cycles under bias load. Furthermore a TWSME is observed in the response of the material under the stress-free thermal loading cycles as depicted by the blue curve. The strain-temperature response for this material, predicted by the proposed model, is shown in figure \ref{fig:TWSME_Simulation}. The predictions are in good agreement with the experimental results. The model successfully captured the evolution on the materials response by predicting adequately the evolution of TRIP strain while it also captured the TWSME of the material using the internal stress term.
It is demonstrated that the proposed model can be used as an efficient tool to predict the evolution of the materials response of untrained SMAs and therefore to enable the design and analysis of SMA-based actuators for real engineering applications.


\vspace{-0.5cm}

\newpage
\section{CONCLUSIONS}\label{conc}
This work presents a three-dimensional constitutive model for the martensitic phase transformation of untrained SMAs subject to thermomechanical cyclic loading conditions. The model is able to predict the traditional thermomechanically induced phase transformations in SMAs. Moreover, by introducing a set of additional internal state variables, i.e., TRIP strain tensor and internal stress tensor, the model is further demonstrated to account for the evolution of irrecoverable TRIP strains and the TWSME for untrained SMAs after cyclic loading. The capabilities of the proposed model are demonstrated by comparing its predictions with experimental results. In particular, boundary value problems considering untrained NiTi SMAs subjected to isothermal and isobaric cyclic loading conditions were investigated. It was shown in the isothermal case that the proposed model was able to predict the generation and saturation of TRIP strains, and also to predict the decreasing stress levels required to start the phase transformation. As for the isobaric loading case, it was demonstrated that the proposed model was able to capture the evolution of materials response during the cyclic loading and to reproduce the TWSME for trained SMAs at stress-free condition. 

The proposed model is anticipated to be further validated against additional experimental data of NiTi and NiTiHf SMAs under uniaxial and other non-uniform loading conditions. The ultimate objective is to validate the capability of the proposed model to predict the response of SMA-based actuators, such as SMA beams and torque tubes, which are intended to be integrated with the future supersonic transport aircrafts to realize the morphing capabilities to reduce the sonic boom noise.


%

\section{Acknowledgments}\label{Ack}
This work is supported by the National Aeronautics and Space Administration (NASA) through the University Leadership Initiative (ULI) project under the grant number: NNX17AJ96A. The conclusions in this work are solely made by the authors and do not necessarily represent the
perspectives of NASA.

\bibliographystyle{aiaa} 
\bibliography{myarticle}

\end{document}